**Second harmonic generation and X-ray diffraction studies of pretransitional region in relaxor $K_{(1-x)} Li_xTaO_3$**

Hiroko Yokota and Yoshiaki Uesu

Department of Physics, Faculty of Science and Engineering, Waseda University,

3-4-1 Okubo, Shinjuku-ku, Tokyo 169-8555, Japan

Charlotte Malibert[1] and Jean-Michel Kiat[1,2]

[1]Structures, Proprietes, Modelisation des Solides, UMR 8580 au CNRS, Ecole Centrale Paris, Grande Voie des Vignes, 92295 Chatenay Malabry Cedex, France

[2]Laboratoire Leon-Brillouin, CE Sacray, 91191Gif-sur-Yvette Cedex, France

Optical second-harmonic-generation(SHG) observations and precise X-ray diffraction experiments have been performed on quantum paraelectrics $KTaO_3$ (KTO) and relaxor $K_{(1-x)}Li_xTaO_3$ with $x$=3% (KLT-3) and 7% (KLT-7). It is found in KLT-3 and KLT-7 that a pretransitional region exists between two characteristic temperatures $T_d$ and $T_p$ ($<T_d$). The average symmetry of the region is tetragonal with a weak lattice-deformation but non-polar in average. The temperature interval between $T_d$ and $T_p$ is consistent with the interval on which neutron diffuse scatterings have been previously reported. These facts strongly suggest that polar micro-regions (PMRs) nucleate around $T_d$ and grow toward $T_p$. Below $T_p$, a larger deformation and a field-induced SH intensity start to develop, while no significant SHG appear in zero-field cooling process. The temperature dependence of the SH intensity below $T_p$ coincides well with that of the tetragonality determined from the lattice deformation. The Landau-Devonshire phenomenological approach suggests that the ferroelectric phase transition at $T_p$ is of first order and that it approaches the second order transition with the decrease of Li concentration. A marked increase of neutron diffraction intensities below $T_p$ indicates that PMRs are transformed to ferroelectric micro-domains at $T_p$, and the micro-domains change to macroscopic ones under the electric field below $T_p$.

# 1. INTRODUCTION

Quantum paraelectrics are dielectrics in which the condensation of a low-lying transverse optical mode is prevented by the quantum fluctuation of atoms. As a result, the divergence of the dielectric constant indicating a ferroelectric phase transition does not occur down to 0K, although the dielectric constant increases with decreasing temperature toward a high value of several thousands. In this sense, quantum paraelectrics are incipient ferroelectrics and a small addition of impurities or an application of external forces can induce ferroelectricity. In particular, with the substitution of Li ions in quantum paraelectric $KTaO_3$(KTO) (henceforth, we use an abbreviation KLT-$x$% for $K_{(1-x)}Li_xTaO_3$), a dielectric peak appears at low temperature and the peak temperature ($T_m$) shifts to higher with the Li concentration.[1] However, the polar state below $T_m$ is not well understood. Some reports insist that it is a long-range ferroelectric phase,[2,3] and others explain it with the dipole glass picture.[1,4] X-ray diffraction study reveals that a tetragonal strain appears above $x$=5%, which indicates that the long-range order develops with higher Li concentrations.[5] On the other hand, Toulouse et al. pointed out that KLT is a kind of relaxor based upon dielectric dispersion characters[6] and neutron diffuse scattering experiments.[7] Recent observations using an optical second harmonic generation(SHG) microscope also support the relaxor nature of KLT-3 because of the marked history dependence of SHG intensities[8] similar to a prototype relaxor $Pb(Mg_{1/3}Nb_{2/3})O_3$(PMN).[9, 10] To be a real relaxor, following criteria should be satisfied; (1) the characteristic frequency dispersion obeying the Vogel-Fulcher law and (2) the appearance of polar micro-regions (PMRs) in a temperature range above $T_m$. Additionally, the history dependence and the slow time evolution of the order parameter are significant indications of relaxors. Although some of these characters have been reported for specific Li concentrations, the characteristic temperature (Burns temperature $T_d$) where PMRs appear has not explicitly discussed yet. For this purpose, two kinds of measurements are necessary, one being sensitive to a lattice deformation and another to a polar state. With this motivation, we investigate KLTs using the SHG microscope which is quite susceptible to polarization changes, and X-ray diffraction to minute lattice deformations. In addition to these measurements, neutron scattering experiments and dielectric measurements are performed on KLT-3 and KLT-7, and pure KTO as a reference crystal to prove the above-mentioned criteria.

## 2. EXPERIMENTAL CONDITIONS

A pure KTO single crystal is grown by the top-seeded crystal growth method. KLT-3 and KLT-7 single crystals are grown by the self-flux method with $Ta_2O_5$, $Li_2CO_3$ and an excess of $K_2CO_3$ as a flux.[11] The exact Li concentration $x$ is determined by the empirical relation between $x$ and the transition temperature $T_P$,[12] $T_P$ being defined from the disappearance temperature of SH intensity in zero-field heating(ZFH) after field-cooling (FC) process. See the details in ref.8.

SHG images are obtained by the SHG microscope, which provides the distribution of the SH intensities in a specimen.[13,14] We use a Q switched $Nd^{3+}$:YAG (yttrium aluminum garnet) laser with the wavelength of 1064 nm, the repetition frequency of 20 Hz, the energy per pulse of 15mJ as a fundamental light source. The polarization direction can be changed by a half-wave plate. SH waves with the 532 nm wavelength generated in a specimen are detected by an image-intensified charge coupled device (ICCD) camera, located behind an analyzer and a band-path filter. The typical exposure time is about 10 sec and is changed from 5 to 15 sec depending on the magnitude of the SH intensity. A specimen is put in a liquid He cryostat (CF2101, Oxford Instruments) for microscopy. Temperature stability of the specimen during the exposure time is about 0.5K.

(100) plate specimens are cut from single crystals and optically finished. Au electrodes are evaporated on a (100) top surface and an electric field is applied along the [001] direction in the sample plane. The incident laser beam of the diameter of 3mm is positioned between two electrodes so as to avoid a photo-conductivity observed at low temperature.

X-ray diffraction measurements are performed on a highly accurate two-axis diffractometer with a Bragg-Brentano geometry using Cu-$K_{\beta}$ wavelength ($\lambda$=1.39223Å) from a 18kW rotating anode generator, equipped with a liquid He cryostat. The lattice constants are determined from profiles of (400) reflections by using the pseudo Voigt analysis. We use (001) plate specimens with the area of 5x3 $mm^2$ and the thickness of 0.65 mm.

Neutron diffraction experiments are performed using a triple-axis spectrometer T1-1HQR installed in JRR-3M in Japan Atomic Energy Agency with the incident energy of neutron beams of 13.60meV ($\lambda$ = 2.459Å). The horizontal collimator sequence is monochromator-20'-sample-40'-detector. Sample dimensions are 6.0x5.5x3 $mm^3$ for KLT-3 and 4.5x4.5x3 $mm^3$ for KLT-7.

The complex impedance $Z'$ (real part) and $Z''$ (imaginary part) of the sample are measured by Solartron SI-1255B frequency response analyzer with SI-1296 dielectric interface. The complex dielectric constant $\varepsilon'$ (real part) and $\varepsilon''$ (imaginary part) are calculated from the impedance and sample dimensions. Measurements of temperature dependence are performed from 20K to 290K with the frequency range from 10 Hz to 1MHz with ac amplitude of 1 $V_{rms}$.

## 3. EXPERIMENTAL RESULTS AND ANALYSES

### A. SHG microscopic observations

Temperature dependences of SH intensity distributions of pure KTO, KLT-3 and KLT-7 under the electric field *E* of 80 V/mm are observed using the SHG microscope. The observations are carried out in different paths in the *E*-*T* diagram; i.e., zero-field cooling(ZFC), ZFH after ZFC(ZFH/ZFC), FC, field-heating after FC (FH/FC) and ZFH after FC(ZFH/FC) processes. Among them, ZFC and ZFH/ZFC processes give no significant SHG for all specimens, and ZFH/FC and FH/FC show almost the same results within experimental errors. Fig.1 shows temperature dependences of SH intensity averaged over homogeneous area (0.05mm$^2$) in SHG images in FC and FH/FC processes. SH intensities of pure KTO at 22K are almost 100 times weaker than other 2 specimens, and the latter exhibits significant SHG signals increasing with the Li concentration. As shown in the SHG images, distributions of SH intensity are inhomogeneous due mainly to the fact that a small portion of paraelectric phase still exists. Stronger electric field (>>100V/mm) would increase the portion of polar regions, but we do not apply it because of the high photoconductivity observed below 70K.

Weak SH intensities of pure KTO are observed only below 30 K, while much stronger SH intensities of KLT-3 and KLT-7 appear below 50 K and 90K, respectively. This phenomenon indicates a macroscopic polar phase appears at the temperatures ($T_p$) under the electric field. Noticeable thermal hysteresis of about 7 K is observed in KLT-7, but not in KLT-3 within temperature fluctuation. Contrary to FH/FC process, FH/ZFC shows quite different features as shown in Fig.2 for KLT-7. Weak SH intensities at low temperature start to increase with raising temperature, shows a maximum then decreases and vanishes at 90K. Similar phenomenon is also observed in KLT-3,[8] and is caused by a domain evolution characterized by fast and extremely slow relaxation times.[15,16]

### B. X-ray diffraction studies

Absolute values of lattice constants of pure KTO, KLT-3 and KLT-7 at room temperature are determined using several (h00) reflections. The result is shown in Fig.3 with the previously reported values[17] for comparison. A small difference of 0.01% is observed in the absolute values but the relative variations with the Li concentration are consistent with the reported values. The lattice constant *a*(*x*) of KLT-*x* decreases almost linearly with *x*. Assuming that the lattice contraction is caused by an effective ionic radius $R_0(Li^+)$ of $Li^+$, we obtain the following relation,

$$a(x) = a(pureKTO)\left\{1 + \frac{\Delta R}{R(K^+)} \times x\right\}, \quad (1)$$

where $R(K^+)$ is an ionic radius of $K^+$ and

$$\Delta R = R_0(Li^+) - R(K^+). \quad (2)$$

From the experiment, we obtain

$$\frac{\Delta R}{R(K^+)} \sim -0.0048. \quad (3)$$

When we take into account the fact that Li ions occupy off-center positions with deviations δ from the ideal A sites, the effective radius of Li ion is expressed as $\delta + \{R(Li^+)/2\}$. From the present experiment and using the ionic radii of $K^+$ (1.64Å for the coordination number of 12) and $Li^+$(0.59~0.92Å depending on the coordination number),[18] δ is estimated to be 1.17~1.34Å, which coincides well with the value of 1.2Å obtained by the NMR measurement.[13] This implies that the deviation δ is equally oriented along the <100> direction in $KTaO_3$ rigid lattices of KLT-3 and KLT-7.

Temperature dependences of lattice constants of KTO, KLT-3 and KLT-7 are plotted in Fig.4. These are measured in ZFC process. We also make the measurements in FC and FH/FC processes. However, as a remarkable photoconductivity rises with X-ray illumination below 70K, an efficacious electric field can not be applied.

As shown in Fig.4, the temperature dependence of the lattice constant of pure KTO is well fitted by the Debye formula,

$$a = a_0 + A \cdot 9RT \cdot \left( \frac{T}{\theta_D} \right)^3 \int_0^{\theta_D / T} \frac{t^3}{e^t - 1} dt \quad , \tag{4}$$

with parameters $a_0 = 3.98317(3)$ Å, $A = 10^{-6}$ Åmol/J, $\theta_D = 312(16)$ K.

In KLT-3 and KLT-7, the splitting of the (400) spectrum as a result of tetragonal deformation is observed at low temperature and becomes remarkable with the Li concentration. Although the result agrees qualitatively with the reported values,[5,8] a substantial difference exists: The present experiment using the high-precision diffractometer reveals for the first time that the tetragonal deformation develops in 2 steps. With decreasing temperature, a small deformation appears at $T_d$ (90K for KLT-3, and 140K for KLT-7) prior to a larger deformation at $T_p$ (50K for KLT-3, and 90K for KLT-7).

To make the fact clearer, the tetragonality $t$ defined by $2(c - a)/(c + a)$ is plotted together with SH intensities in FH/FC process in Fig.5. It should be noted that the electric-field induced SHG appears at $T_p$ and not at $T_d$. This fact strongly suggests that a pretransitional region exists between $T_d$ and $T_p$. The region is accompanied with a small tetragonal deformation but non-polar in average because of a lack of SHG. Below $T_p$, the temperature dependence of the tetragonality coincides well with that of SH intensity under an electric field.

The SH intensity $I^{(2\omega)}$ is expressed by the following expression.[19]

$$I^{(2\omega)} = 2\omega^2 \left( \frac{\mu_0}{\varepsilon_0} \right)^{\frac{3}{2}} \frac{1}{n^{(2\omega)} \left\{ n^{(\omega)} \right\}^2} \left( \varepsilon_0 d \right)^2 \{ I^{(\omega)} \}^2 \frac{4 \sin^2 \left( \frac{1}{2} \Delta k L \right)}{\left( \Delta k \right)^2} . \quad (5)$$

Here ω indicates the frequency of the fundamental wave, $n^{(\omega)}$ and $n^{(2\omega)}$ the refractive index of the fundamental and SH waves, respectively, $I^{(\omega)}$ the intensity of the fundamental wave, $\Delta k$ the misfit parameter of wave numbers between the fundamental and SH waves, and $L$ the specimen thickness. The constant $d$ is a SHG tensor component, and in the present experiment, it is $d_{33}$ in the Voigt notation with the polarization directions of the fundamental and SH waves parallel to the [001] direction. Since the product of $d_{33}$ and $P_3$ is invariant upon the group theoretical consideration,[20] $d_{33}$ changes linearly with $P$. Therefore $I^{(2\omega)}$ is proportional to $P^2$. Through the electro-strictive effect, the tetragonality $t$ is also proportional to $P^2$. This explains the coincidence between $I^{(2\omega)}$ and $t$ observed below $T_p$.

### C. Phenomenological analyses of experimental results

We apply the Landau-Devonshire phenomenological theory to the temperature dependences of SH intensities in FH/FC processes. The free energy expanded up to the sixth power of $P$ is expressed as [21]

$$F = \frac{1}{2}\alpha (T - T_0) P^2 + \frac{1}{4}\beta P^4 + \frac{1}{6}\gamma P^6 . \quad (6)$$

The equilibrium and stability conditions give [22]

$$P^2 = \frac{-\beta + \sqrt{\beta^2 - 4\alpha\gamma(T - T_0)}}{2\gamma}. \qquad (7)$$

The experimental results of SHG are fitted using Eq.(7) and shown in Fig.6. In the figure, the experimental result of KLT-1.7 under the same magnitude of $E$ is also shown. The coefficients in the free energy determined by the fittings are tabulated in Table I, where the magnitudes of α of KLT-3 and KLT-7 are fixed at the values estimated from the dielectric constants in higher temperature region. The calculated curves agree well with the experimental ones. It should be noted that the order of phase transition of KLT-1.7 is the second since $\beta>0$ and the first for KLT-3 and KLT-7 ($\beta<0$) and that the first-order nature becomes stronger with the increase of the Li concentration. In fact, the maximum hysteresis width $\Delta T_{hys} = \beta^2/4\alpha\gamma$ of KLT-7 calculated with the free-energy coefficients is 3K, while much smaller value of $\Delta T_{hys}$ (= 0.014K) is obtained for KLT-3. The result is also supported by the experimental fact that a clear thermal hysteresis is found in KTL-7 but not in KLT-1.7 and KLT-3.

### D. Dielectric measurements

Temperature dependences of dielectric constants of pure KTO and KTO with doped with different Li concentrations have already been reported.[6,23-25] The essential feature of our results is same as those reported. Here we show our results of KLT-3 and KLT-7 to clarify the behaviors around both characteristic temperatures $T_d$ and $T_p$, and to compare the parameters determined using the Vogel-Fulcher law with those of PMN.[26]

Fig.7 shows the result of KLT-3 in ZFH/ZFC, with the real part ε' in (a), and the imaginary ε" in (b). In addition to a pronounced peak with the characteristic frequency dispersion, small peaks are observed around $T_d$ (=90K), in particular, in the imaginary part. As shown in an inserted figure of Fig.7(b), $T_m$ shifts to higher side with increasing frequency. This kind of dispersion is characteristic in heterogeneous dielectrics and could be related to the appearance of PMRs.

Fig.8 shows the result of KLT-7 in ZFH/ZFC process, where (a) indicates ε', and (b) ε". The dispersion around $T_d$ (=130K) is not clearly observed in comparison with KLT-3, probably because the strong first order nature at $T_p$ hides the small changes around $T_d$. In lower temperature, a new feature becomes visible: A sudden drop of ε' is observed at $T_p$=90K. This phenomenon was also reported in KLT-3.5 in lower frequency region,[6] however, a steep change in ε' is more clearly observed in higher frequency region. The reason of this difference is not clear at present. In any way, this shows that the ferroelectric phase transition takes place at $T_p$ in some portions of the specimen. We shall discuss it more precisely in the next session.

Observed dielectric responses are analyzed with the Vogel-Fulcher law connecting the frequency $f_m$ and the temperature $T_m$ of the maximum ε' in a dipole glass or a spin glass system:

$$f_m = f_0 \exp[-E_a / k_B(T_m - T_{VF})] \quad , \qquad (8)$$

where $f_0$ means the Debye frequency, $E_a$ the activation energy, $k_B$ the Boltzmann factor and $T_{VF}$ is termed the freezing temperature below which dipoles cannot respond to the electric field. Table II shows the parameters of KTL-3, KTL-7 determined by the present experiments and those of PMN.[26] The activation energy of KLT-7 is larger than that of KLT-3. This fact would be related to the increase of the anisotropy energy with the Li concentration, since $E_a$ is the product of an electro-crystalline anisotropy energy density $K_{anis}$ and the volume of PMR. [26] On the other hand、no marked difference is found in the Debye frequency $f_0$ between KLT-3 and KLT-7, while these values are one order magnitude smaller than that of PMN, which could be due to restrained motions of Li ions locating strong off-center positions.

It should be noted that $T_{VF}$ in relaxors is different from the freezing temperature in a dipole glass or a spin glass, because of much more complicated nature of the dipolar relaxation.[27] In fact, the response of dipole moments of KLT-3 under the electric field exhibits a sudden slowing-down at 40K, which is higher than $T_{VF}$.[16]

### E. Neutron scattering experiments

Neutron elastic scattering experiments are carried out on KLT-3 and KLT-7. Fig.9(a) and (b) show temperature dependences of (220), (200) reflection intensities of KLT-3 and KLT-7, respectively. Both crystals exhibit steep increases of the intensities below $T_p$. This phenomenon can be attributed to the change of mosaicity of the crystal due to the appearance of ferroelectric micro-domains which develop below $T_p$. The mosaicity of crystal breaks the extinction law and makes the Bragg intensities increase.[28] In KLT-7, a thermal hysteresis of about 7K is observed as shown in Fig.9(b), which agrees with the SHG measurement.

## 4. DISCUSSIONS

The most important finding of the present studies is the existence of a pretransional state between the non-polar cubic phase and the polar tetragonal phase. This fact can be disclosed mainly due to the combination study using a polarity-sensitive SHG microscope and a strain-sensitive X-ray diffraction on the specimens cut from a same single crystal. The possible picture of the pretransitional region is following: In cubic phase, dipole moments produced by off-centered Li ions align along one of the symmetry equivalent <100> directions. Below $T_d$, individual Li dipoles start to correlate to form polar micro-regions (PMRs). Because the polarization is confined in a local area and the orientation is at random even under an electric field, no SHG is observed. On further cooling down to $T_p$, PMRs interact each other with a finite correlation length which becomes a macroscopic scale and strong SH waves are generated in FC process. On the other hand, the lattice strains are induced by the average of polarization fluctuation $<(\Delta P)^2>$ through the electrostrictive effect. Lattice strains $x_1$, $x_3$ and tetragonality $t$ are expressed as

$$x_1 = Q_{13}<(\Delta P)^2>, \quad x_3 = Q_{33}<(\Delta P)^2>, \text{ and } \quad t = x_3 - x_1 = (Q_{33} - Q_{31})P_s^2 \quad (9).$$

This explains the quite small lattice anisotropies ($t$ = $2x10^{-4}$ for KLT-3 and $4x10^{-4}$ for KLT-7 in the vicinity of $T_p$) in the pretransitional region. Below $T_p$, the same expressions should be hold with the same electrostrictive constants $Q_{13}$ and $Q_{33}$ but $\Delta P$ is replaced by the spontaneous polarization $P_s$. Since the magnitude of $P_s$ is much larger than that of $\Delta P$, larger strains and consequently larger tetragonality are induced as shown in Fig.5.

We would point out here that the structural phase transition at $T_d$ from $Pm\bar{3}m$ to nonpolar tetragonal phase, i.e., $P4/mmm$ with the long-range order, seems to be inconceivable, because previous studies using infrared,[29] and hyper Raman[30] did not detect any change in phonon anomalies near $T_d$. We also performed neutron inelastic scattering experiments using pure KTO, KLT-3 and KLT-7, and found that low-lying TO and TA phonons do not show any anomalies around $T_d$. These evidences would support that $T_d$ is the Burns temperature and not a phase transition temperature with the long-range order.

The phase below $T_p$ is ferroelectric even without the external electric field. It consists of ferroelectric micro-domains, which are supported by the neutron elastic scattering experiments described in the previous section. Sizes of micro-domains without an electric field are estimated to be less than the wavelength of the used laser, as appreciable SHG cannot be observed without an electric field.

The picture of PMR in the intermediate phase explain well the similarity between KLT and prototype relaxor PMN, e.g., characteristic dielectric dispersion obeying the Vogel-Fulcher Law, the history dependence of $P$ under an electric field, the slow kinetics of $P$ under $E$, or non-ergodic behavior of it. Another experimental support is the agreement of the temperature interval $\Delta T$ of the pretransitional region where the neutron diffuse scatterings are observed:[7] In KLT-6, $\Delta T$ is estimated to be 30~40K, which is consistent with our results.

Based upon the idea, the polarization states in 3 phases in KLT are schematically illustrated as shown in Fig.10. In ZFC process, the high temperature phase is non-polar cubic $Pm\overline{3}m$ with uncorrelated individual shifts of Li and below $T_d$, PMRs appear in the cubic matrix. In the vicinity of $T_d$, the density of PMR is not large and each polar region fluctuates independently to form a super-paraelectric state. On approaching to $T_p$, PMRs start to interact each other to grow the area with size-distribution, which provides characteristic dielectric dispersion. Below $T_p$, a ferroelectric phase transition takes place accompanied by the micro-domain state.

In FC process, the polarization direction is easily aligned below $T_p$ and a macroscopic polarization can be produced and maintained down to the low temperature region. In ZFC process at sufficiently low temperature, a random force originated from the random distribution of Li dipoles freezes a polar state, but the electric field cannot change the polarization direction. Thus in FH/ZFC process, a macroscopic polarization cannot be produced under the electric field up to a temperature near 40K where the thermal excitation overcomes the random force.

Finally we would stress that KLT-3 and KLT-7 are belonging to a new class of perovskite relaxors where the A-sites are occupied by different kinds of isovalent atoms. Because almost all relaxors have a structure with B-sites occupied by 2 kinds of heterovalent ions, KLT will be an important prototype relaxor for understanding the nature of relaxors.

This work is supported by the grants-in-aid of scientific research (A) of MEXT, "Academic Frontier" Project of MEXT, and the $21^{st}$ century COE program “Physics of systems with self-organization composed of multi-elements” of MEXT, Japan. We are grateful to Y.Yamada and Y.Tsunoda in Waseda Univ., M.Maglione in Institute of Condensed Matter Chemistry of Bordeaux, S.Vakhruchev in Ioffe Physico-Technical Institute, and R.E.Cohen in Carnegie Institution of Washington, for their invaluable discussions.

Table I. Coefficients in the Landau-Devonshire theory

| Substance | $T_0$ [K] | α | β | γ |
|---|---|---|---|---|
| KLT-1.7 | 34.5 | $1.0x10^{-5}$ | $3.20x10^{-1}$ | $4.53x10^{3}$ |
| KLT-3 | 47.3 | $1.13x10^{-5}$ | $-1.49x10^{-2}$ | $3.44x10^{2}$ |
| KLT-7 | 85.9 | $1.16 x10^{-5}$ | $-1.76x10^{-1}$ | $2.16x10^{2}$ |

Table II. Experimentally determined parameters in the Vogel-Fulcher law of KLT-3, KLT-10 and PMN.

| Substance | $f_o$ [Hz] | $E_a$ [eV] | $T_{VF}$ [K] | |
|---|---|---|---|---|
| KLT-3 | $9.7 \times 10^9$ | 0.045 | 0.045 | Present study |
| KLT-7 | $4.8 \times 10^{10}$ | 0.088 | 56.3 | Present study |
| PMN | $1.03 \times 10^{12}$ | 0.0407 | 291.5 | ref.25 |

**References**

* Corresponding author. Electronic address: hiroko-9bq@ruri.waseda.jp

**Figure Captions**

Fig.1 Temperature dependences of SH intensities of KLT-3 (a) and KLT-7 (b) in FH/FC process. SHG images taken at some temperatures are also shown. In (b), the open triangles indicate the result of FC process, and the solid squares that of FH/FC.

Fig.2 Temperature dependence of SH intensities of KLT-7 in FH/ZFC (solid squares) and FH/FC processes (gray circles).

Fig.3 Dependence of the absolute lattice constant $a$ of KLT-$x$ on Li concentration $x$.

The solid squares indicate the present results and the open triangles those reported in ref.17.

Fig.4 Temperature dependences of lattice constants of KTO (open squares), KLT-3 (gray triangles) and KLT-7(solid circles). The Debye fitting for KTO is plotted with the dotted line.

Fig.5 Temperature dependences of tetragonalities of KLT-3 (solid squares) and KLT-7 (solid triangles) in ZFC process and SHG intensities KLT-3 (open squares) and KLT-7 (open triangles) in FH/FC process.

Fig.6 Temperature dependences of SH intensities KLT-1.7, KLT-3 and KLT-7 in FH/FC process. The solid lines are the fitting curves using the Landau-Devonshire theory.

Fig.7 Temperature dependences of real part ε' (a) and imaginary part ε" (b) of the complex dielectric constant of KLT-3 as a parameter of frequency.

Fig.8 Temperature dependences of real part ε' (a) and imaginary part ε" (b) of the complex dielectric constant of KLT-7 as a parameter of frequency.

Fig.9 Temperature dependences of (220) neutron Bragg reflection intensities of KLT-3 (a), and (200) intensities of KLT-7 (b).

Fig.10 Schematic illustration of the polarization states in 3 phases in KLT. Gray region indicates non-polar state and black and gray regions polar state with different

polarization orientations.

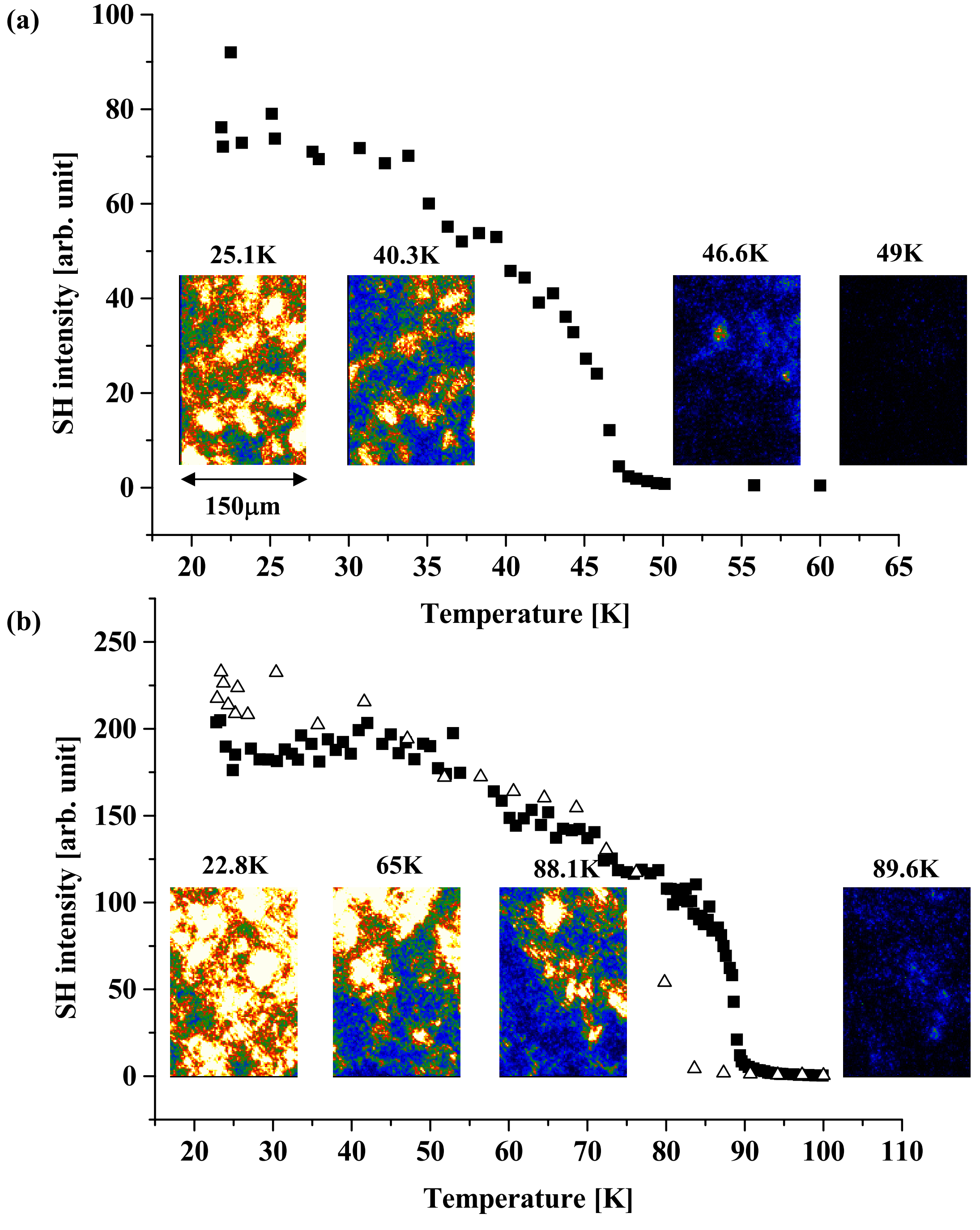

2 Columns in Colors

Fig.1 H. Yokota et al, PRB

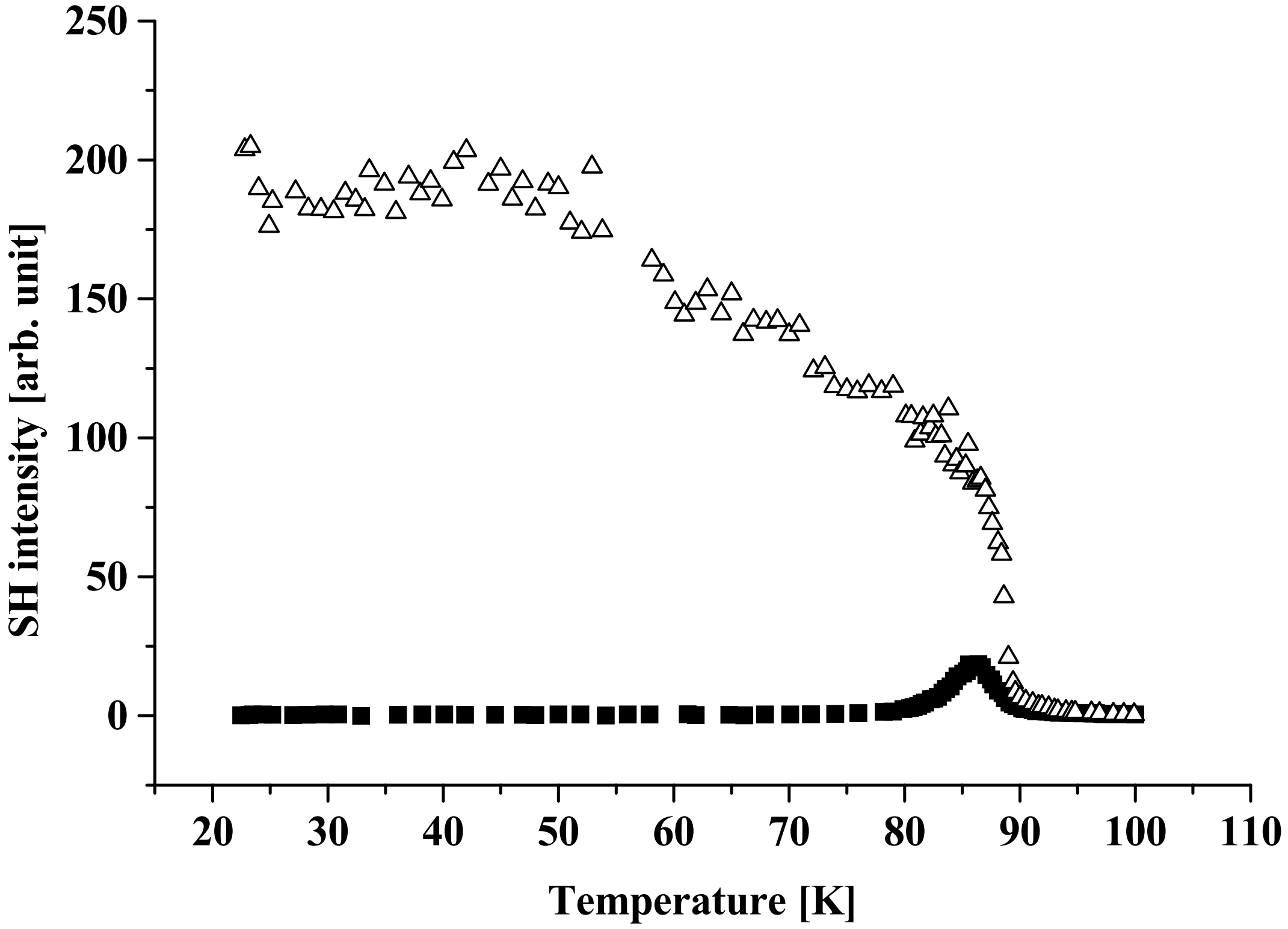

**Fig.2 H. Yokota et al, PRB**

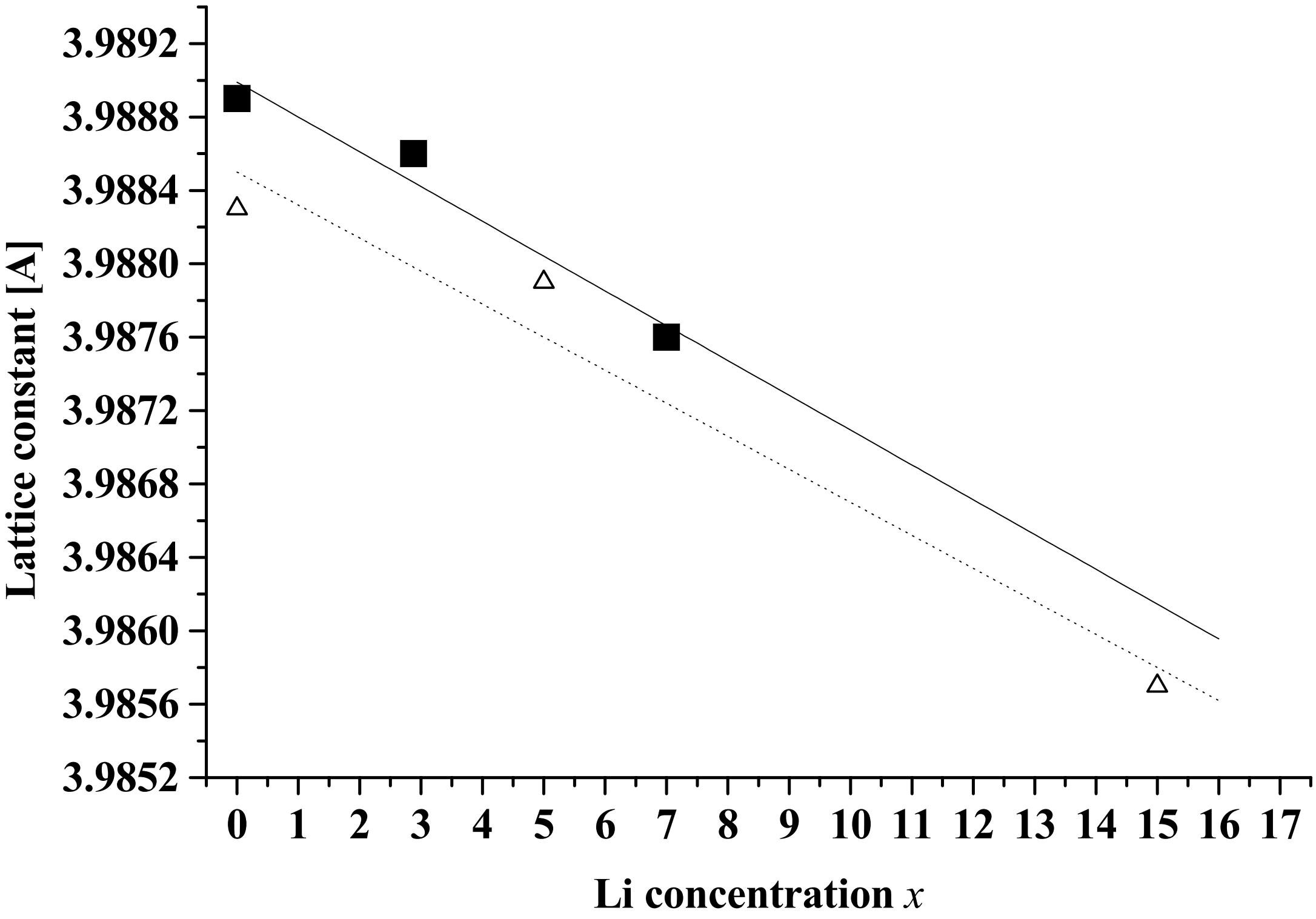

**Fig.3 H. Yokota et al, PRB**

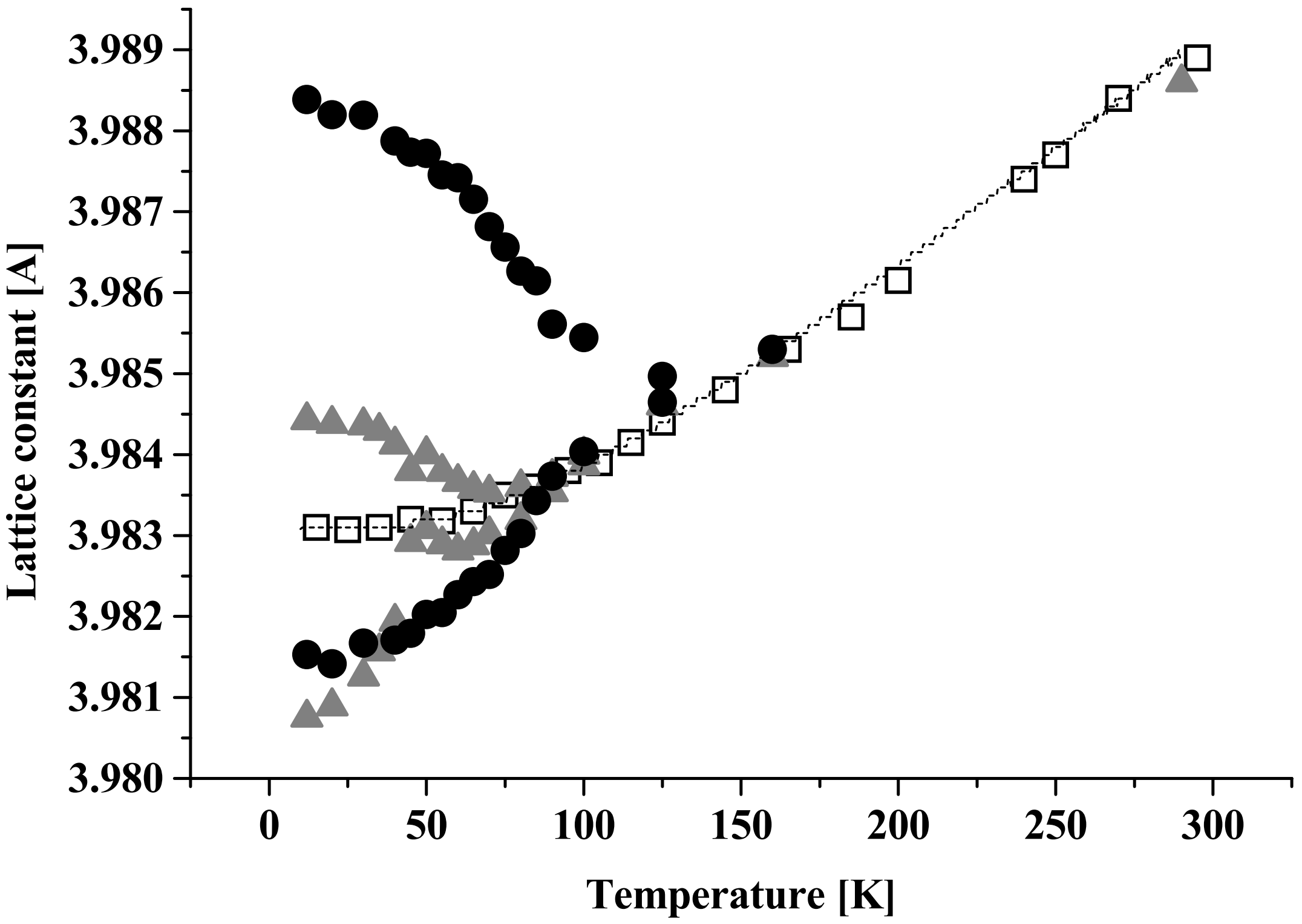

**Fig.4 H. Yokota et al,PRB**

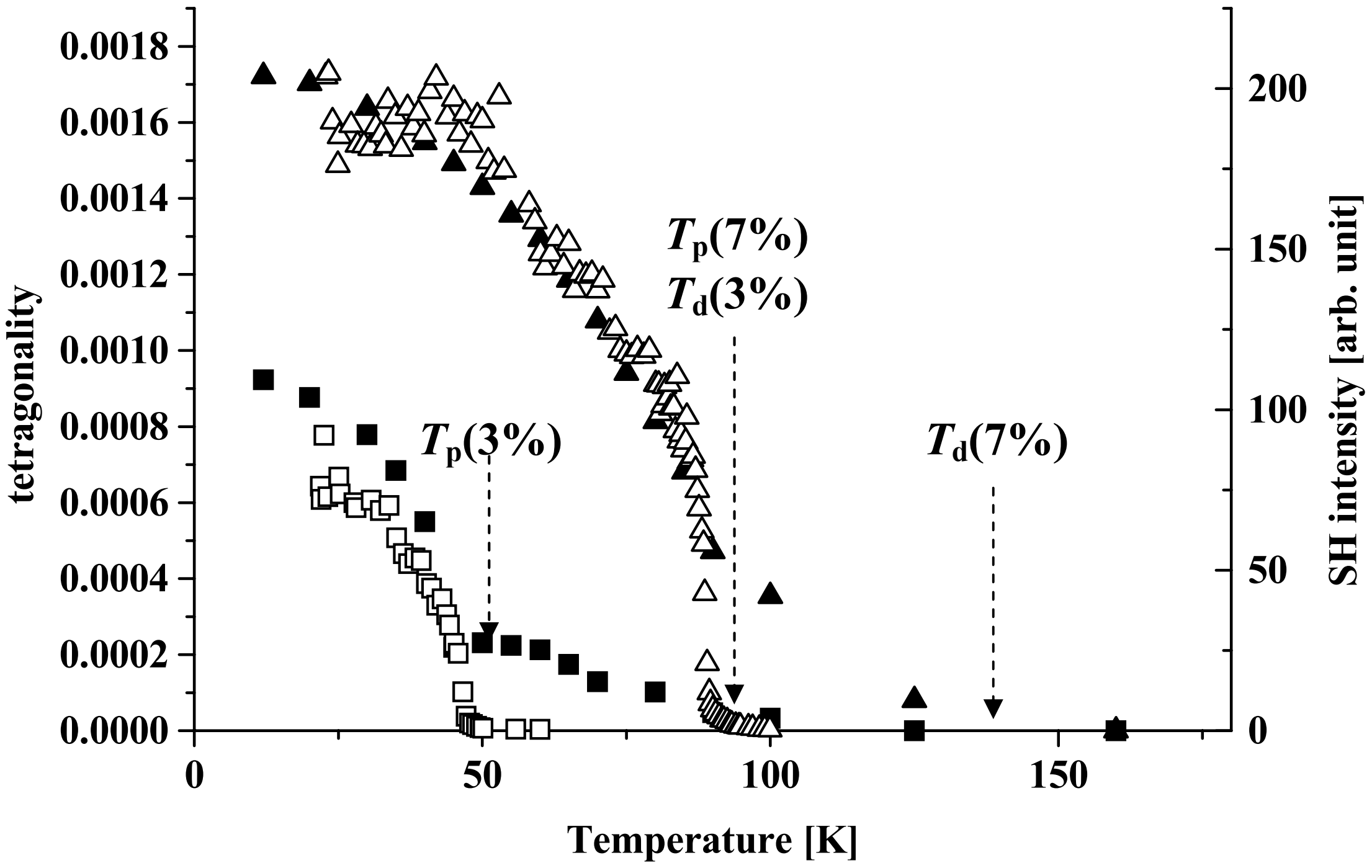

Fig.5 H. Yokota et al, PRB

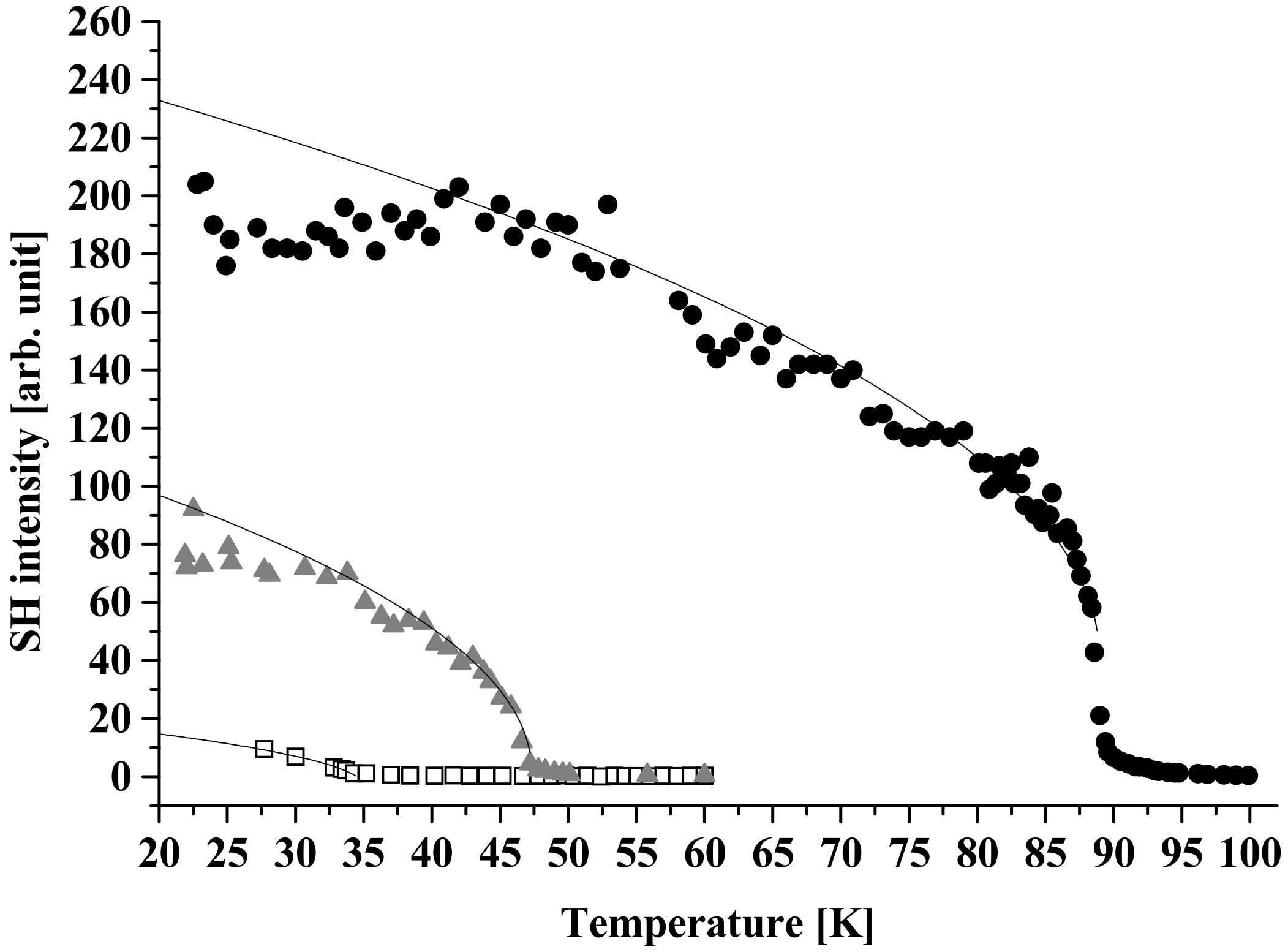

**2 Columuns**

**Fig.6 H. Yokota et al, PRB**

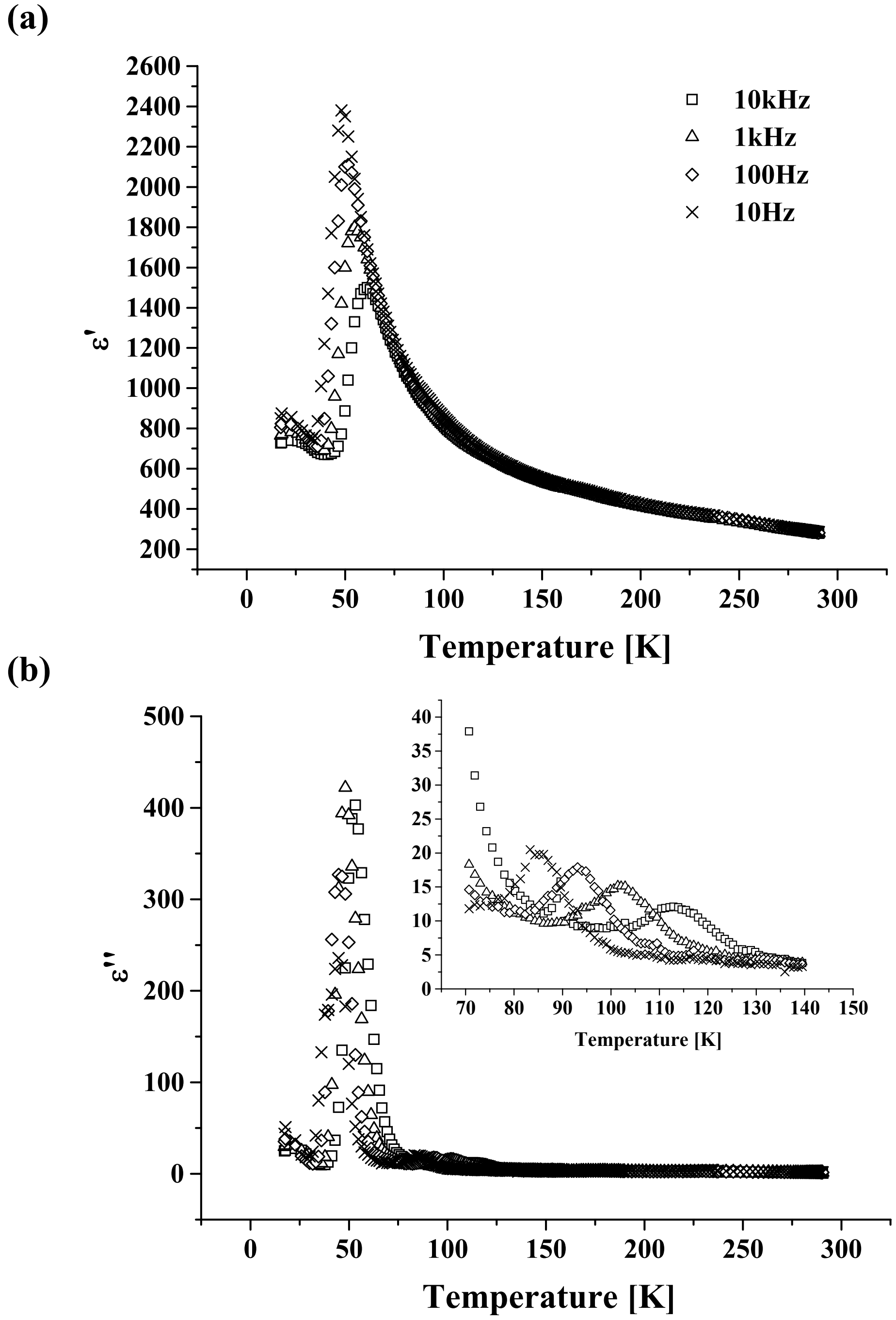

**Fig.7 H. Yokota et al, PRB**

(a)

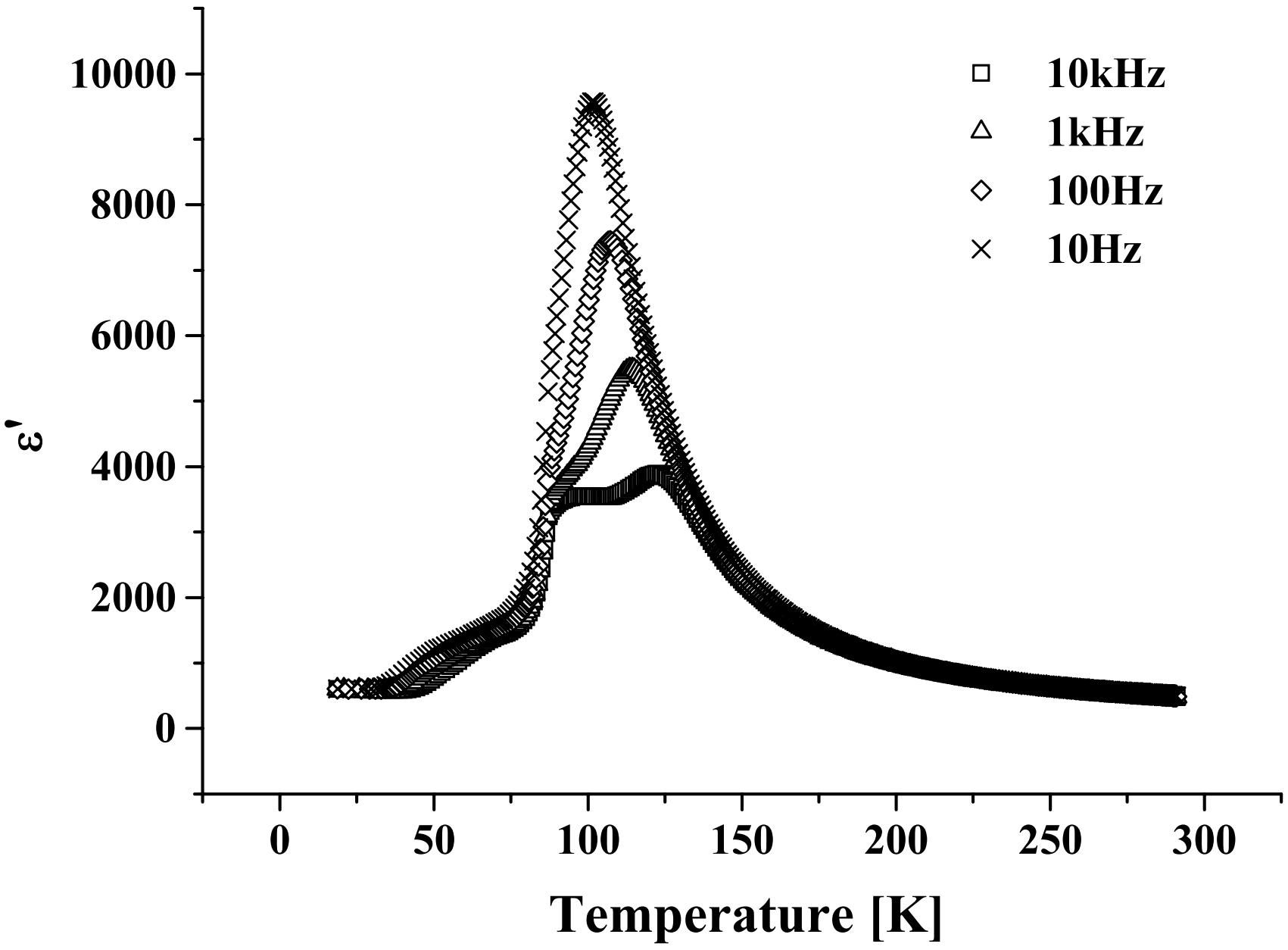

(b)

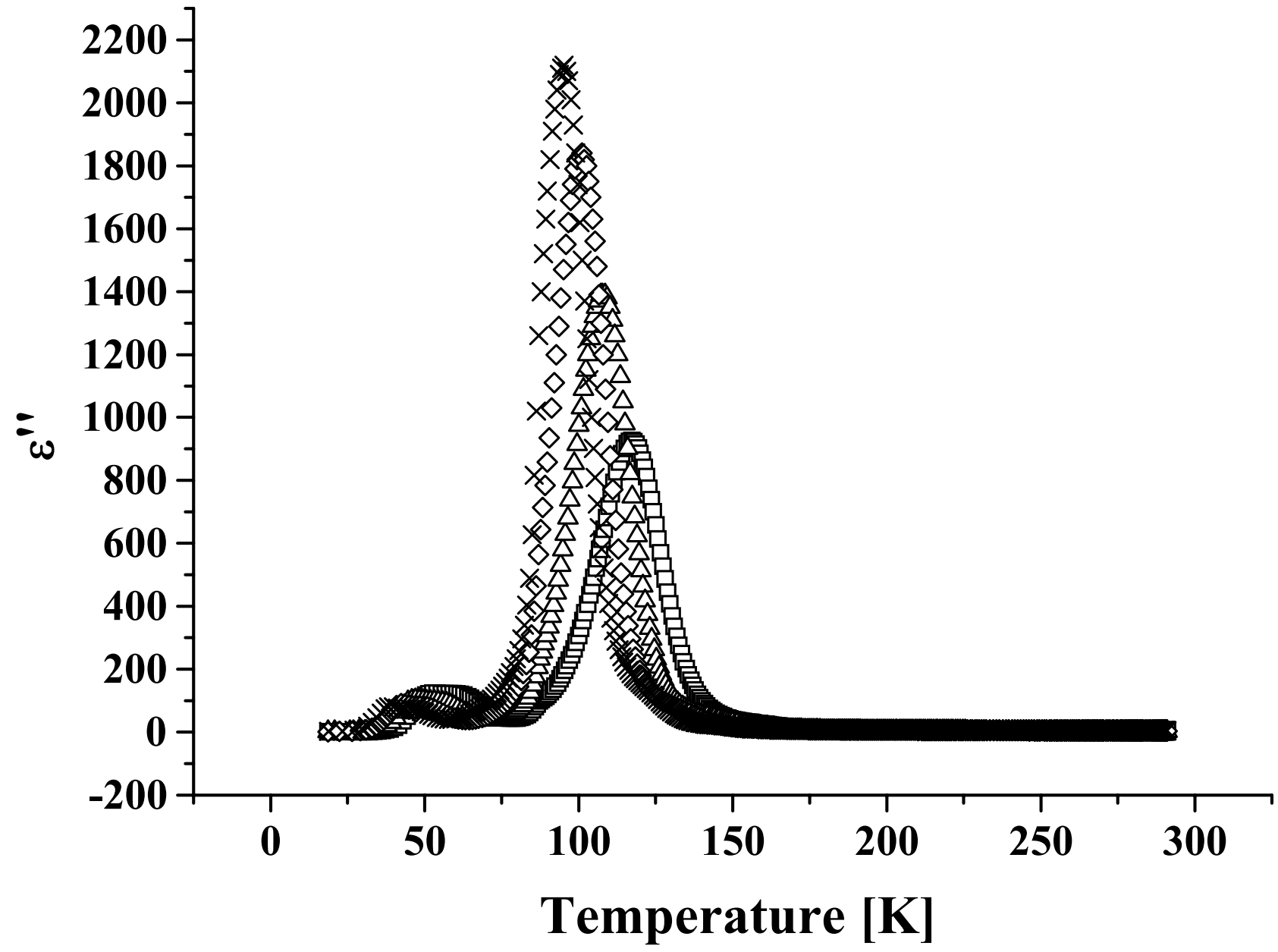

**Fig. 8(b) H. Yokota et al, PRB**

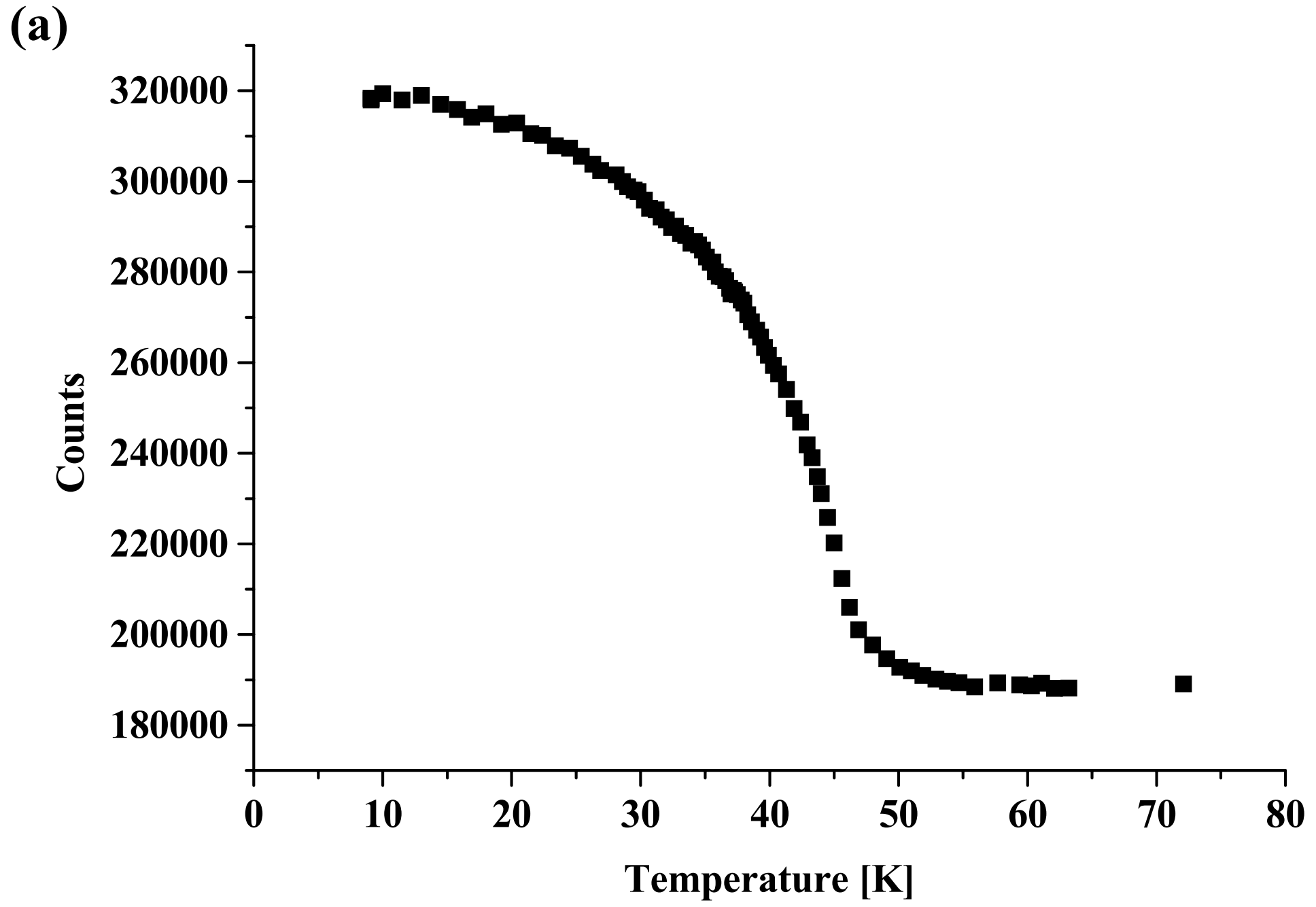

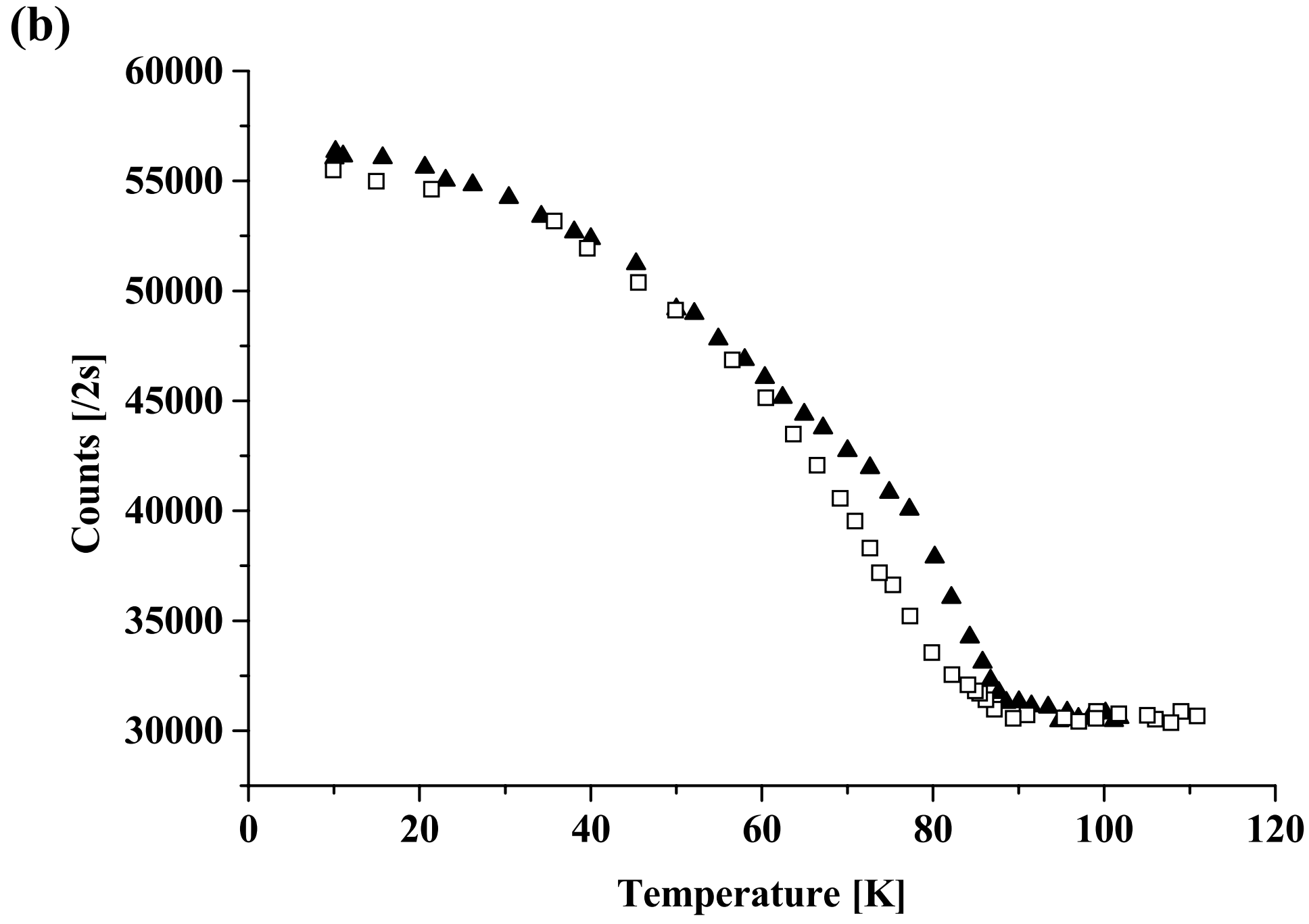

**Fig. 9 H. Yokota et al, PRB**

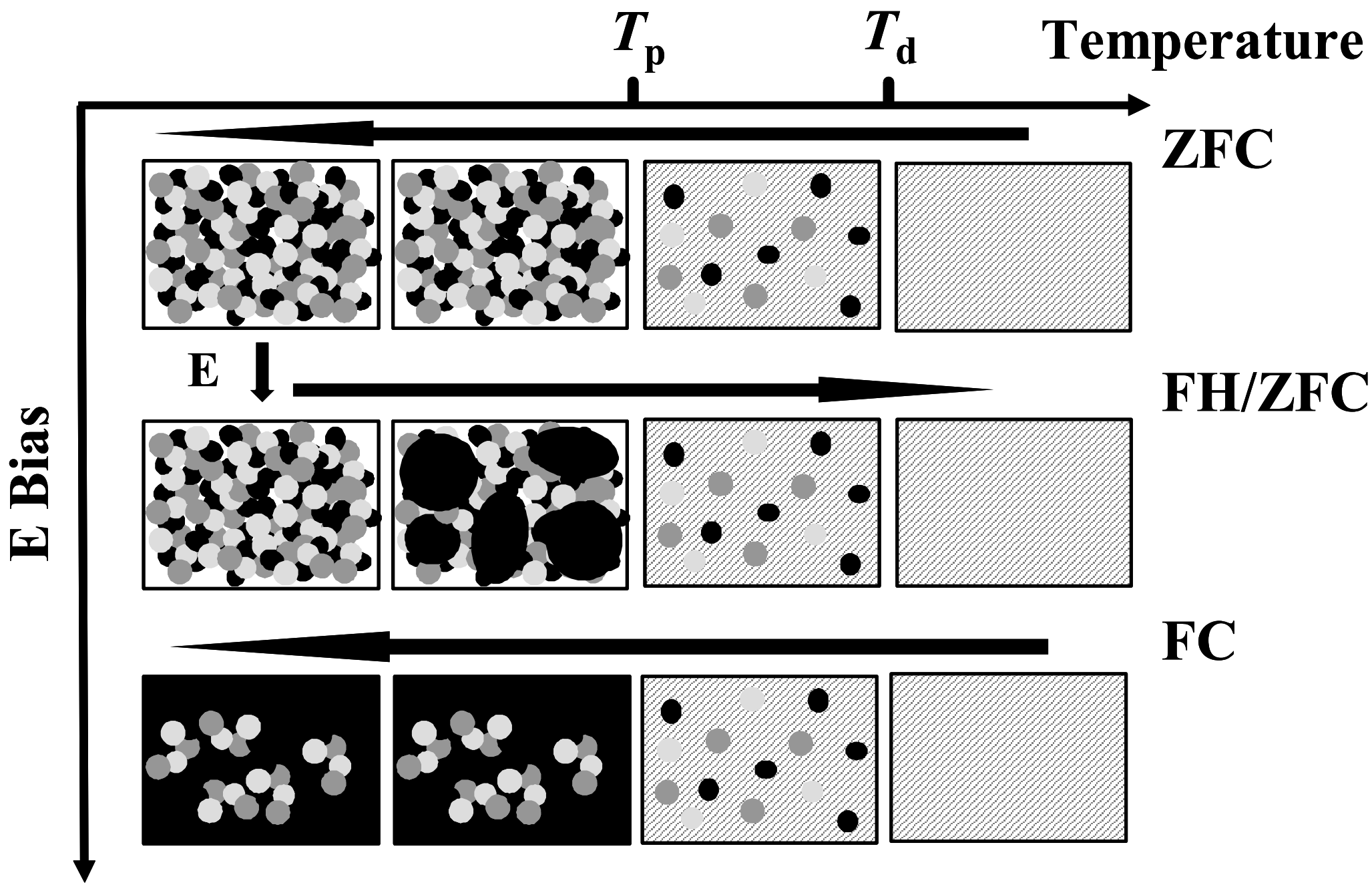

Fig.10 H.Yokota et al, PRB